\documentclass[%
 reprint,
showpacs,preprintnumbers,
footinbib,
 amsmath,amssymb,
 aps,
pra,
floatfix,
]{revtex4-1}

\usepackage[T1]{fontenc}
\usepackage[utf8]{inputenc}
\usepackage{amsmath}
\usepackage{ mathrsfs }
\usepackage{amsthm}
\usepackage{amsfonts,dsfont}
\usepackage{graphicx}
\usepackage{bbm}
\usepackage{hyperref}
\usepackage{color}
\usepackage{comment}

\theoremstyle{theorem}

\newcommand{\ket}[1]{| #1 \rangle } 
\newcommand{\upd}{\mathrm{d}}
\newcommand{\tr}{\mathrm{tr}}
\newcommand{\ie}[0]{\textit{i.e.} }
\newcommand{\eg}[0]{\textit{e.g.} }

\definecolor{cbl}{rgb}{0,0,1}

\definecolor{crd}{rgb}{1,0,0}

\begin{document}
\title{Efficient progressive readout of a register of (qu)bits}
\author{Antoine Tilloy}
\email{antoine.tilloy@ens.fr}
\affiliation{Laboratoire de Physique Th\'eorique, Ecole Normale Sup\'erieure de Paris (PSL), France
}

\date{\today}

\begin{abstract}
Recently, a series of articles by Combes \textit{et al.} has shown that it was possible to greatly improve the measurement rate of a register of qubits for given detector resources by means of a clever feedback control scheme. However, this speed-up came at an exponential cost in terms of complexity and memory use. In this article, I propose a simple efficient algorithm --exponentially more frugal in memory and less complex to implement-- which is asymptotically as fast. I use extensively the implicit classicality of the situation to provide a slightly more straightforward interpretation of the results. I compute the speed-up rates exactly in the case of the proposed model and in the case of the open-loop scheme of Combes \textit{et al.} and prove that they indeed provide the same asymptotic speed-up.
\end{abstract}
\maketitle

\section{Introduction} 

Measuring a quantum system usually takes a non negligible amount of time. In some cases, this time turns out to be much larger than the typical timescales of the system dynamics, making \eg system characterization and measurement-based control difficult. In the future, finite measurement times may also put constraints on the performance of quantum computers by limiting the speed at which large qubit register can be read out. Procedures that can reduce this measurement time while using the same detector resources are thus interesting both from a theoretical and practical point of view. 

In a recent series of articles \cite{combes2008,combes2010,combes2011,combes2015}, Combes \textit{et al.} have proposed control schemes which increase the measurement speed of qubit registers. The methods they proposed provide an impressive speedup rate proportional to the register size. However, in contrast to the simple no control procedure, they require a prohibitive exponential amount of memory. In addition to their relative complexity, this latter limitation makes these new procedures implementable only for small qubit registers. It may have seemed that the use of an exponential complexity was the price to pay for this linear speed-up: ``\textit{You cannot have your cake and eat it}''. Fortunately, it turns out that this is not the case here. In this article we introduce a \emph{quasi} open-loop scheme that gives a similar gain while requiring much fewer control operations on the system ($\simeq 2$ on average for typical parameter values) and using only a linear amount of memory.

This article is structured as follows. We shall first introduce briefly in section \ref{sec:measurement} the quantum trajectory formalism for the continuous measurement of a qubit and show that, at least in the case we consider, it is formally equivalent to a fully classical probabilistic model. We shall then review in section \ref{sec:rapidmeasurement} the previous approaches to rapid measurement before presenting our own model and deriving its properties in section \ref{sec:model}. We will then briefly review additional numerical results in \ref{sec:numerics} and conclude by discussing possible improvements and extensions. The proof that the schemes provide the claimed speed-up rates are rather cumbersome and relegated to appendices.

\section{Continuous measurement of a register of (qu)bits}\label{sec:measurement}

The standard way to describe the progressive measurement of a quantum system is by means of repeated interaction schemes. A quantum system is coupled briefly with a ancilla which is subsequently measured. As the interaction has entangled the two quantum systems, measuring the ancilla gives some information on the system of interest. The measured ancilla is then discarded and a new one is sent to interact with the system before being measured again. Iterating this procedure many times then gives a progressive measurement of the system in a basis which is fixed by the system-ancilla unitary interaction. In the limit where this procedure is repeated infinitely frequently with an infinitely small interaction time, one gets a continuous stochastic evolution for the system which is called a quantum trajectory \cite{brun2002,jacobs2006}. In what follows we will first give, without proof, the equations one gets for a continuously monitored qubit. Then we will show that, in the specific case we consider, the same equations can be derived from a much simpler classical model. This classical picture provides valuable insights and the reader unfamiliar with quantum trajectories is encouraged to take it as the starting point.

Using standard continuous quantum measurement theory \cite{barchielli1986,caves1987,diosi1988,barchielli1991,wiseman1996,belavkin1992,wiseman2009}, one can show that a qubit of density matrix $\rho\in \mathds{C}^2\otimes\mathds{C}^2$ subjected to the continuous measurement of the operator $\sigma_z$ obeys the Stochastic Master Equation (SME):
\begin{equation}\label{eq:qubitsme}
\upd \rho_t = 2 \gamma \mathcal{D}[\sigma_z](\rho_t) \,\upd t + \sqrt{2\gamma} \mathcal{H}[\sigma_z](\rho_t)\,\upd W_t
\end{equation}
where $\gamma$ codes for the measurement strength, $W_t$ is a Wiener process and we have used the standard notation from \cite{wiseman2009}:
\begin{equation}
\begin{split}
\mathcal{D}[A](\rho)&=A\rho A^\dagger -\frac{1}{2}(A^\dagger A \rho + \rho A^\dagger A)\\
\mathcal{H}[A](\rho)&=A\rho + \rho A^\dagger - \tr\left[ (A+A^\dagger)\rho\right]\rho
\end{split}
\end{equation}
The associated measurement signal, which is the continuous and weak equivalent of the (random) result from a Von Neumann measurement, reads:
\begin{equation}\label{eq:signal}
\upd Y_t = 2\sqrt{2\gamma} \, \tr (\sigma_z \rho_t)\, \upd t + \upd W_t
\end{equation}
In the absence of proper Hamiltonian for the qubit, it is easy to notice that the phases of the density matrix in the eigenbasis of $\sigma_z$ are exponentially suppressed and have no back-action on the diagonal coefficients. Consequently, we can consider, without lack of generality, that we start from a diagonal density matrix. In that case, eq. \eqref{eq:qubitsme} takes the simple form:
\begin{equation}\label{eq:simplifiedsme}
\upd p_t = 2\sqrt{2\gamma} \,p_t (1-p_t)\, \upd W_t
\end{equation}
where $p_t$ is simply the probability to be in the state $\ket{0}=\ket{+}_z$ at time $t$: $p_t = \langle0|\rho_t|0\rangle$. An interesting feature of this equation is that it is completely classical. In the absence of phases, eq. \eqref{eq:qubitsme} can be understood simply as the fancy quantum rewriting of an inherently classical model where a classical bit has a well defined value which is progressively revealed. Let us make this claim more precise by explicitly constructing the equivalent classical model.

Consider a classical bit that can take two values $R=0$ and $R=1$ (or equivalently $+$ and $-$). The bit state is fixed but unknown. An observer progressively extracts information about the bit state by doing a succession of imperfect classical measurements with results $\delta_k=\pm 1$. A measurement gives some, but not all, the information on the system state, more precisely we take:
\begin{equation}
\begin{split}
&\mathds{P}(\delta_k=1 | R=0)=\frac{1+\varepsilon}{2}\\
&\mathds{P}(\delta_k=1 | R=1)=\frac{1-\varepsilon}{2}
\end{split}
\label{eq:def}
\end{equation}
which fully specifies how the information is extracted.
The parameter $\varepsilon \in \;]0,1[$ codes for the quality of the measurement and we will be interested in the very bad measurement limit $\varepsilon \rightarrow 0$. We write $\mathcal{F}_k=\sigma\left(\left\{ \delta_i\right\} , i\leq k\right)$ the natural filtration associated to the stochastic process of the measurement results. In other words, $\mathcal{F}_k$ corresponds to the intuitive notion of the information contained in the measurement results up to step $k$. The quantity of interest is the probability $p_k=\mathds{P}\left(R=0|\mathcal{F}_k\right)$, \ie the probability for the bit state to be zero knowing the first $k$ measurement results. A simple application of Bayes' rule gives the following update rule for $p_k$:
\begin{equation}\label{eq:discrete}
p_{k+1}=\frac{(1+\varepsilon\,\delta_{k+1})\,p_k}{(1+2\varepsilon\,\delta_{k+1})(p_k-1/2)}.
\end{equation}
This discrete update rule becomes a set of stochastic differential equations in the appropriate weak measurement limit $t=k \upd t$, $\varepsilon =\sqrt{\gamma\upd t}$ and $Y_t=\sqrt{\upd t} \sum_{i=1}^k \delta_i$ (see \textit{e.g.} \cite{spikes}):
\begin{equation}
\left\lbrace
\begin{split}
\upd p_t &= 2\sqrt{2\gamma} \,p_t (1-p_t)\, \upd W_t\\
\upd Y_t &= 2\sqrt{2\gamma}\left(2p_t-1\right) \, \upd t + \upd W_t
\end{split}\right.
\end{equation}
Which is exactly the same set of equations as in the quantum case. From now on we will thus use the classical picture when it makes the proofs mathematically simpler or just more intuitive. The reader unfamiliar with continuous quantum measurement can also simply keep the previous classical picture in mind and stop being bothered with quantum mechanics, at least regarding the rest of this article.

\begin{figure}
\includegraphics[width=\columnwidth,trim = 0cm 22.5cm 5cm 0cm, clip]{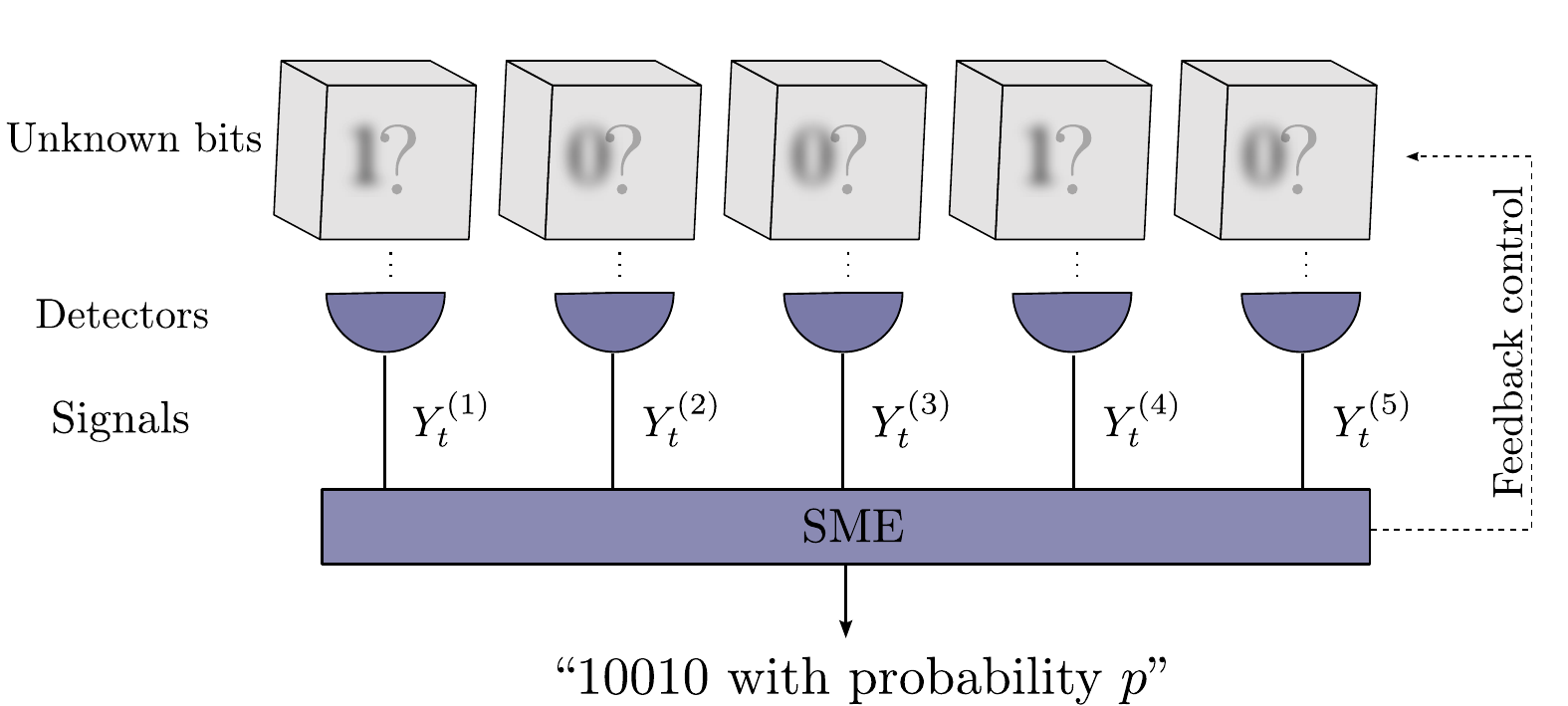}
\caption{Schematics of the continuous measurement of a register of bits with feedback control.}
\label{fig:register}
\end{figure}

Building upon the previous construction, it is easy to describe the continuous measurement of a register of $n$ qubits (which, with the same argument as before, is equivalent to the continuous measurement of a register of classical bits). We assume that all the (qu)bits are measured independently by $n$ detectors, the density matrix $\rho\in \left(\mathds{C}^2\otimes\mathds{C}^2\right)^{\otimes n}$ verifies the SME:

\begin{equation}\label{eq:arraysme}
\upd \rho_t = 2 \gamma\sum_{i=1}^n \mathcal{D}[\sigma_z^{(i)}](\rho_t) \,\upd t + \sqrt{2\gamma}\sum_{i=1}^n \mathcal{H}[\sigma_z^{(i)}](\rho_t)\,\upd W_t^{(i)}
\end{equation}
where:
\begin{equation}
\sigma_z^{(i)} = \mathds{1}\otimes \mathds{1} \otimes ... \otimes \sigma_z \otimes ... \otimes \mathds{1}
\end{equation}
with $\sigma_z$ in $i$-th position. The Wiener processes are uncorrelated, \ie $\upd W_t^{(i)}\upd W_t^{(j)}=\delta_{ij}\upd t$. The signals $Y^{(i)}$ associated to the detectors verify the same equation as before:
\begin{equation}
\upd Y_t^{(i)} = 2\sqrt{2\gamma} \, \tr \left(\sigma_z^{(i)} \rho_t\right)\, \upd t + \upd W_t^{(i)}.
\end{equation}
Everything can be rewritten with the help of a classical vocabulary in the same way as before. Assuming one has no prior information on the \emph{true} state $R\equiv \left(R^{(1)},...,R^{(n)}\right)$ of the classical register, all the bits can be considered independently in the sense that the total probability factorizes:
\begin{equation}
\mathds{P}\left[R|\sigma(\mathcal{F}^{(1)}_t,...,\mathcal{F}^{(n)}_t)\right]=\mathds{P}\left[R^{(1)}|\mathcal{F}^{(1)}_t \right]... \mathds{P}\left[R^{(n)}|\mathcal{F}^{(n)}_t\right].
\end{equation}
It is only necessary to compute the evolution of the probabilities of the $n$ single bit states, or say of the $n$ marginals, to know the probability of a register configuration. Storing the $2^n$ probabilities of all the register configurations is not needed in this simple measurement setup. This helpful property will unfortunately be lost for more elaborate measurement schemes.

To quantify the rate at which information is extracted as a function of time with the continuous measurement scheme, the (now standard) approach is to consider the log-infidelity ln $\Delta$ where $\Delta=1-\lambda_0$ with $\lambda_0$ the largest eigenvalue of $\rho$ (or equivalently here, the probability of the most probable register configuration). In addition to its simplicity, this measure has conceptual advantages which are detailed in \cite{combes2008}. We shall not elaborate on this fact here and simply assume that the log-infidelity is a relevant measure of information extraction. 

Writing the density matrix in a basis where the most probable state is noted $\bar{0}=\ket{00...0}$ and using It\^o rule, one gets after a straightforward computation:
\begin{equation}\label{eq:infidelity}
\mathds{E}\left[\upd \ln\Delta_t\right] = -4\gamma \sum_{i=1}^{n} \tr\left[(\sigma_z^{(i)}-\mathds{1}) \rho_t\right]^2 \frac{(1-\Delta_t)^2}{\Delta_t^2}\upd t
\end{equation}
In the simple case I consider here, it can be shown (see \eg \cite{combes2015}) that for large time, \ie $t\gg\gamma^{-1}$ the previous expression simplifies to:
\begin{equation}
\mathds{E}\left[\upd \ln\Delta_t\right] = -16\gamma\,\upd t
\end{equation}
The objective of rapid measurement schemes is to improve this convergence rate while still using the same detector resources.

\section{Standard rapid measurement schemes}\label{sec:rapidmeasurement}

Before going into the details of the rapid measurement schemes, we should give an intuition of why some asymptotic speed-up is expected. Consider that the previous measurement scheme has been run for a while and look at the two most probable configurations. Being able to discriminate rapidly between these two configurations is what makes a measurement scheme fast, at least in a first approximation. However, because the probability of a configuration can be written as a product of single bit probabilities, the two most probable configurations differ only by one bit. Consequently, only one detector actually provides relevant information while the $n-1$ other ones are essentially useless. Intuitively, one can expect that a good measurement scheme will find a way to harness the information extraction ability of the $n$ detectors at the same time. Doing so should naively provide a speed up of order $n$ (and we will see that this is what all the algorithms get).

Let us now recall what is allowed for a rapid quantum measurement scheme. In contrast with rapid purification schemes \cite{jacobs2003,combes2006,wiseman2006,combes2010pur}, the control applied on the system should commute with the measured observables, \ie the control should not amount to a change of the measurement basis. The only operations that are consequently allowed are permutations of vectors in the measurement basis\cite{combes2008}. Incidentally, this means that the evolution of the density matrix with the control is still equivalent to that of a classical system because the control itself is a purely classical operation; rapid quantum measurement is inherently a classical problem. This classicality is very helpful to understand what the measurement result means after such a procedure. If we stick to the quantum, we have to say that the measurement result allows us to retrodict what the system state would have been in the absence of control. Classically, the system state is fixed, it is then subjected to a measurement procedure with operations that can easily be reversed at the end, once the result is known. Using the mathematical equivalence of the situation, we can thus say that everything happens \emph{as if} the system state were fixed but unknown at the beginning and that the optimal measurement procedure simply revealed it.

Let us now review briefly the first proposal of Combes \textit{et al.} \cite{combes2008} for a locally optimal measurement scheme. A brief look at eq. \eqref{eq:infidelity} shows that the locally optimal case is obtained for a permutation of the initial basis that maximises the quantity:
\begin{equation}
\sum_{i=1}^{n}\tr\left[(\sigma_z^{(i)}-\mathds{1}) \rho_t^{LO}\right]^2
\end{equation}
where $\rho_t^{LO}$ is the optimally permuted density matrix.

The asymptotic speed-up is then defined as:
\begin{equation}
S_{LO}=\lim_{t\rightarrow+\infty}\mathds{E}\left[\frac{(1-\Delta_t)^2}{4\Delta_t^2}\sum_{i=1}^{n} \tr\left[(\sigma_z^{(i)}-\mathds{1}) \rho_t^{LO}\right]^2\right].
\end{equation}
which is the asymptotic ratio of the convergence rates for the locally optimal case and the no feedback case.
Following \cite{combes2008}, the key concept is the Hamming distance \cite{hamming1950}. The Hamming distance between two states counts the number of bit differences between them. The idea is to do a permutation of the pointer basis which puts the next-to-most probable states as far as possible (with respect to the Hamming distance) from the most probable one. This will maximise $\sum_{i=1}^{n}\tr\left[(\sigma_z^{(i)}-\mathds{1}) \rho_t^{LO}\right]^2$. More precisely, one first needs to order all the states but the most probable one by decreasing order of probability in a first list, then to order the states in decreasing order of Hamming distance with respect to the most probable one in a second list. The control then consists in mapping the states of the first list to the states of the second list while keeping the most probable state unchanged. Intuitively, one expects this scheme to provide a speed up of order $n$ because the probable states can be discriminated from the most probable one with approximately $n$ detectors at the same time. And indeed, in \cite{combes2008} the authors manage to prove that for large $n$:
\begin{equation}\label{eq:LO}
\frac{n}{4} \leq S_{L0} \leq n
\end{equation}

The previous locally optimal scheme requires a real time feedback loop which may be difficult to implement in practice. It would be more convenient to have a predefined strategy implementable in open-loop. In \cite{combes2015} the authors provide such a scheme. The idea is simply to do rapid random permutations of the pointer basis. One expects that, on average, the states will be at a Hamming distance of order $n/2$ from each other yielding the same kind of speed-up as before but for a different pre-factor. And indeed, in \cite{combes2015} the authors prove that the speed-up $S_{RP}$ for the open-loop random permutation scheme verifies for large $n$:
\begin{equation}
\frac{n}{4} \leq S_{RP} \leq \frac{n}{2}
\end{equation}
Actually, it is possible to prove that the upper bound is reached exactly, \ie that:
\begin{equation}
S_{RP} = \frac{n}{2}\,\frac{2^n}{2^n-1} \underset{n+\infty}{\sim} \frac{n}{2},
\end{equation}
see appendix \ref{appendix:RP}. This result will allow us to compare this scheme with our new measurement procedure more precisely.

These two schemes are certainly appealing but they suffer from an important limitation which makes them essentially impossible to implement on future large registers.
Setting aside the astonishingly high number of permutations $N_P=\left(2^n\right)!$ that are needed in the open-loop case (because a smaller number, say only exponential, might give similar speed-ups), the main obstacle is that the two schemes require an exponential amount of memory to store the $2^n$ diagonal coefficients of the full density matrix, \ie the probabilities of all the register configurations. Indeed because of the successive permutations of the pointer basis, it is no longer true that the probability of a configuration can be reconstructed from a product of the $n$ marginals, a lot of information is stored in the bit correlations. Additionally, the two schemes require that the operator do a huge number of permutations on the system, something which may be difficult to implement in practice.

\section{\emph{``Guess and Check''} Algorithm}\label{sec:model}

\subsection{Description}
Naively, a good way to build a procedure more frugal in memory would be to reduce the number of different basis used in the open-loop scheme hoping that the states still stay far away from one another on average with respect to the Hamming distance. It turns out that requiring that every state is far away from every other one on average is a very demanding requirement which is only needed for a \emph{truly} open-loop control scheme. Slightly relaxing the open-loop condition, it is possible to construct a quasi open-loop, or as we will call it ``\emph{guess and check}'' (GC) algorithm which is fast and memory efficient. 

The idea goes as follows. Imagine one knows a good candidate for the register state after running the standard measurement scheme for a while. Then two ordered pointer basis are enough to keep the candidate at an average Hamming distance of $n/2$ from every other state. The solution is simply to take an ordered basis $\mathscr{B}$ and then exchange the candidate and its bitwise opposite to get a new basis $\tilde{\mathscr{B}}$. If a given state is close to the candidate $c$  in $\mathscr{B}$ then it will be far from $c$ in $\tilde{\mathscr{B}}$ and \textit{vice versa}. Measuring successively in $\mathscr{B}$ and $\tilde{\mathscr{B}}$ should yield a convergence speed-up of $n/2$ provided the candidate initially chosen was correct. If this is not the case and the most probable state changes during the measurement process, then a crude yet practical solution is simply to start the whole process again and discard all the information acquired before. The key thing to notice is that the time spent in the guessing process and in eliminating wrong guesses is finite on average and has accordingly no impact on the asymptotic speed-up. This will be discussed in more details later, but let us start by presenting the algorithm precisely this time:
\vskip0.5cm
\textbf{\textit{Guess and Check} measurement protocol}
\begin{enumerate}
\item Run the standard measurement protocol until a register configuration, later called the \emph{candidate}, reaches a probability superior to a predefined threshold $p_0$ (\eg $p_0=1/2$).
\item Implement the permutation mapping the initial ordered pointer basis $\mathscr{B}$ to the new ordered pointer basis $\tilde{\mathscr{B}}$ where the candidate and its bitwise opposite are exchanged.
\item Measure in the two basis by applying the permutation and its inverse successively until the target infidelity is reached or until the probability of the candidate becomes negligibly small (say inferior to $p_0^{'} \ll p_0$).
\item In the latter case, start the whole protocol again from the first step and discard all the information previously acquired.
\end{enumerate}

\begin{figure*}
\includegraphics[width=1.6\columnwidth,trim = 0cm 18.9cm 0cm 0cm, clip]{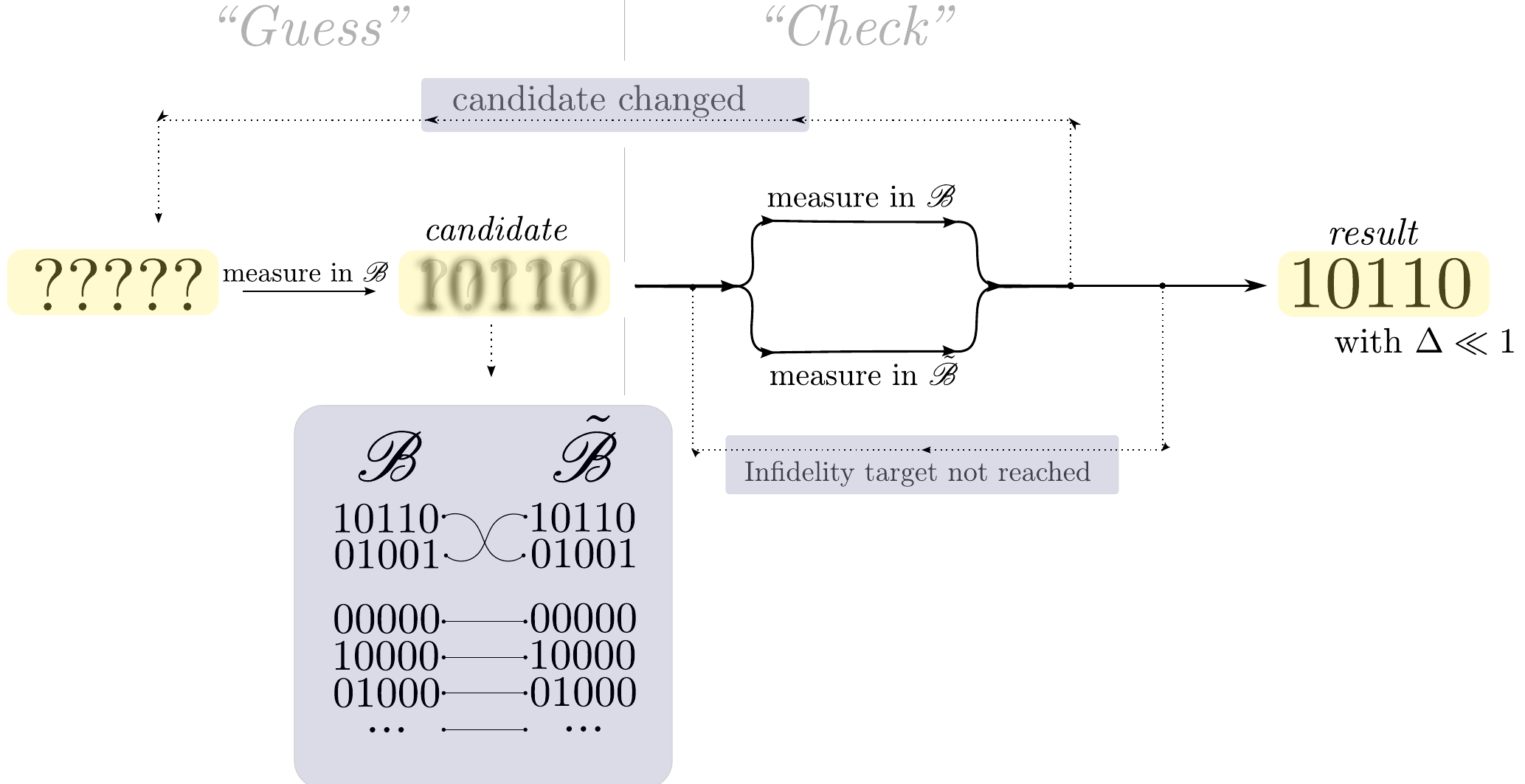}
\caption{Guess and Check algorithm.}
\label{fig}
\end{figure*}

At this point, a few comments are in order. The value of the thresholds, though important for the short term behavior of the protocol, has no impact on the asymptotic behavior of the infidelity. The frequency at which one should switch of measurement basis is voluntarily left open for the simple reason that it does not matter\footnote{Actually the same answer can be given to the program outlined by Combes \textit{et al.} in the conclusion of \cite{combes2015}: ``\textit{Finally and perhaps most importantly [...] future work should include imperfections [...] such as a finite number of permutations in a fixed time interval.}''. As long as all the permutations are sampled in the end, the frequency simply does not matter!}! Indeed, as it is clear from the classical picture, the measurements can be done in any order without changing the statistics (equivalently, one can assume that all the measurement have already been done in the two basis, that the system state is fixed, and that one only progressively reveals the measurement results). The only thing that is needed is that the same amount of time is spent in every basis and that the most probable state is computed from time to time to check if it has changed. Step 4 of the protocol is obviously highly suboptimal and could be improved greatly in the future. However, as it is written now, it has the great advantage of making the whole algorithm very easy to analyse rigorously.

The protocol is \emph{quasi} open loop in the sense that only a small number of actions (finite on average) has to be done by the controller which makes it much easier to implement than a real-time feedback loop. Moreover, the feedback part consists in a simple unitary operation on a 2-level system consisting of the candidate and its bitwise opposite. The protocol is thus simple and robust, the only thing that remains is to show that it indeed provides a speedup of order $n$ and that it only requires a memory of size $\mathcal{O}(n)$.

In what follows and for notational simplicity, we will assume that the candidate is labelled in the same way in $\mathscr{B}$ and $\tilde{\mathscr{B}}$ \ie we will use a \emph{notation} where all the bits are flipped in $\tilde{\mathscr{B}}$. Alternatively, one can consider that $\tilde{\mathscr{B}}$ is obtained from $\mathscr{B}$ via a full bit flip of all the states but the candidate and its bitwise opposite which is strictly equivalent.

\subsection{Speedup of order $n$}
Provided the candidate turns out to stay the most probable state during the whole process, it can be shown (see appendix \ref{appendix:GC}) that the infidelity decreases at a rate $n/2$ times larger than with the standard measurement scheme. More precisely, it can be shown that for large time:
\begin{equation}
\mathds{E}[-\upd \ln \Delta_t]\underset{t+\infty}{\sim} \frac{n}{2} \times 16 \gamma t.
\end{equation}
However, this does not straightforwardly give the speedup rate as the candidate might just turn out not to stay the most probable state forever if the initial guess was wrong. Two additional contributions need to be taken into account. First, some time $\tau_0$ (finite on average) is needed to find a first candidate with probability superior to the predefined threshold. Second, some time $\tilde{\tau}_1$ may be needed to realize that this was not the good candidate. In such a situation, which happens with probability $(1-p_0)/(1-p_0^{'})\simeq (1-p_0) $, we then have to start over and wait for a time $\tau_1$ before finding a new good candidate, which may turn out to be wrong after a time $\tilde{\tau}_2$, and so on. As a result the average time $T$ spent out of the fast converging phase of the algorithm is:
\begin{equation}
T\simeq\mathds{E}[\tau_0] + \sum_{i=1}^{+\infty} (1-p_0)^i \left(\mathds{E}[\tilde{\tau}_i+\tau_i]\right)
\end{equation}
Because we start over every time the candidate turns out to be incorrect, $\tau_i$ and $\tilde{\tau}_i$ are random variables with a law independent on $i$. Consequently, the latter equation reduces to:
\begin{equation}
T\simeq\mathds{E}[\tau] +\frac{1-p_0}{p_0}\left(\mathds{E}[\tau] + \mathds{E}[\tilde{\tau}]\right) 
\end{equation}
which is finite. As a result, the amount of time wasted trying to find the candidate and eliminating incorrect ones does not matter for the asymptotic properties of the log-infidelity. Finally we have, for the \textit{guess and check} algorithm and $n\geq 2$:
\begin{equation}
S_{GC} = \frac{n}{2}
\end{equation}
This means that the \textit{guess and check} procedure offers an asymptotic speedup equal to that of the true open-loop scheme for large $n$.

\subsection{Computation with $\mathcal{O}(n)$ memory}
To prove that we do not need to store the full density matrix, it is easier to use the mathematically equivalent classical picture.
In what follows, we will decompose the information coming from the $2n$ measurement records ($n$ measurement apparatus in two distinct basis). For that matter, it is convenient to introduce the notations:
\begin{equation}
\begin{split}
\mathcal{F}_t^k&=\sigma\left(\left\{ Y^{(k)}_u\right\} , u\leq t\right)\\
\tilde{\mathcal{F}}_t^k&=\sigma\left(\left\{ \tilde{Y}^{(k)}_u\right\} , u\leq t\right)\\
\mathcal{G}_t&=\sigma\left(\left\{Y^{(k)}_u, \tilde{Y}^{(k)}_u\right\} , u\leq t,k=1..n\right)
\end{split}
\end{equation}
where $Y^{(k)}$ is the signal from detector $k$ in $\mathscr{B}$, where $\tilde{Y}^{(k)}$ is the signal from detector $k$ in $\tilde{\mathscr{B}}$ and where $\mathcal{G}_t$ thus contains all the information available up to time $t$. Marginals with respect to the filtrations $\mathcal{F}_t^k$ and $\tilde{\mathcal{F}}_t^k$ can be computed in real time using only one signal and independently of the rest via eq. \eqref{eq:simplifiedsme}. The objective is now to express probabilities with respect to the full filtration as a function of these easily computable marginals. Elementary applications of Bayes' rule give:

\begin{equation}
\mathds{P}[R=s|\mathcal{G}_t]=\frac{1}{Z_t}\prod_{k=1}^{n}\frac{\mathds{P}[R=s|\mathcal{F}^{(k)}_t]\mathds{P}[R=s|\tilde{\mathcal{F}}^{(k)}_t]}{\mathds{P}[R|\mathcal{G}_0]^2}
\end{equation}
where $Z_t$ is the normalization. Assuming equal probability at initial time for simplicity we get:
\begin{equation}
\mathds{P}[s|\mathcal{G}_t]\!=\!\frac{1}{\mathcal{Z}_t}\prod_{k=1}^{n}\mathds{P}\!\left[R^{(k)}=s^{(k)}|\mathcal{F}^{(k)}_t\right]\mathds{P}\!\left[\tilde{R}^{(k)}=\tilde{s}^{(k)}|\tilde{\mathcal{F}}^{(k)}_t\right]
\end{equation}
where, again, $R^{(k)}$ (resp. $\tilde{R}^{(k)}$) is the value of the $k$-th bit of $R$ in $\mathscr{B}$ (resp. in $\tilde{\mathscr{B}}$). The difficulty is now that the normalisation factor $\mathcal{Z}_t$ contains an exponential number of terms so that it would seem that we still need an exponential number of operations. However, because of the simple permutation between the two basis, the normalization factor $\mathcal{Z}_t$ can be computed exactly:
\begin{equation}\label{eq:partitionfunction}
\begin{split}
\mathcal{Z}_t=&\sum_{s\in\mathscr{S}} \prod_{k=1}^{n}\mathds{P}\left[R^{(k)}=s^{(k)}|\mathcal{F}^{(k)}_t\right]\mathds{P}\left[\tilde{R}^{(k)}=\tilde{s}^{(k)}|\tilde{\mathcal{F}}^{(k)}_t\right]\\
=&\sum_{s\in\mathscr{S}} \prod_{k=1}^{n}\mathds{P}\!\!\left[R^{(k)}=s^{(k)}|\mathcal{F}^{(k)}_t\right]\!\mathds{P}\!\!\left[\tilde{R}^{(k)}=1-s^{(k)}|\tilde{\mathcal{F}}^{(k)}_t\right]\\
&+\prod_{k=1}^{n} \mathds{P}\left[R^{(k)}=0|\mathcal{F}^{(k)}_t\right]\mathds{P}\left[\tilde{R}^{(k)}=0|\tilde{\mathcal{F}}^{(k)}_t\right]\\
&+\prod_{k=1}^{n} \mathds{P}\left[R^{(k)}=1|\mathcal{F}^{(k)}_t\right]\mathds{P}\left[\tilde{R}^{(k)}=1|\tilde{\mathcal{F}}^{(k)}_t\right]\\
&-\prod_{k=1}^{n} \mathds{P}\left[R^{(k)}=0|\mathcal{F}^{(k)}_t\right]\mathds{P}\left[\tilde{R}^{(k)}=1|\tilde{\mathcal{F}}^{(k)}_t\right]\\
&-\prod_{k=1}^{n} \mathds{P}\left[R^{(k)}=1|\mathcal{F}^{(k)}_t\right]\mathds{P}\left[\tilde{R}^{(k)}=0|\tilde{\mathcal{F}}^{(k)}_t\right], 
\end{split}
\end{equation}
where we have used the fact that for all the states $s$ but two (the candidate $\bar{0}$ and its bitwise opposite $\bar{1}=\ket{11...1}$), $\tilde{s}^{(k)}=1-s^{(k)}$. To simplify the expressions we introduce the compact notations for the marginals in the two basis knowing only the information from one series of measurements: $p^{(k)}_t=\mathds{P}[R^{(k)}=0|\mathcal{F}_t^{(k)}]$ and $\tilde{p}^{(k)}_t=\mathds{P}[\tilde{R}^{(k)}=0|\tilde{\mathcal{F}}_t^{(k)}]$ (which, it should be emphasized, are not the \emph{true} marginals, \ie the marginals conditioned on all the available information). Theses marginals can be computed independently and easily in real time from the measurement records $Y_t^{(k)}$ and $\tilde{Y}_t^{(k)}$ using eq. \eqref{eq:signal} and \eqref{eq:simplifiedsme} (or in the discrete case eq. \eqref{eq:discrete}) . The sum in eq. \eqref{eq:partitionfunction} can be done separately on each term of the product which gives:
\begin{equation}
\begin{split}
\mathcal{Z}_t=&\prod_{k=1}^{n} \left[p_t^{(k)} (1-\tilde{p}_t^{(k)}) +(1- p_t^{(k)}) \tilde{p}_t^{(k)}\right]\\
&+ \prod_{k=1}^{n} (2p_t^{(k)}-1)(2\tilde{p}_t^{(k)}-1)
\end{split}
\end{equation}
where the first term comes from the sum over all states and the second comes from the $4$ correction terms of eq. \eqref{eq:partitionfunction}. This means that the normalization factor can be computed knowing only the $2n$ marginals in the two basis and doing \emph{only} a number $\mathcal{O}(n)$ of elementary operations (additions and multiplications) on them. Eventually, the probability of any state can be computed from a linear number of operations on the marginals. For example, we can compute the probability of the candidate $\lambda_0$:
\begin{equation}
\lambda_0=\frac{p_t^{(k)} \tilde{p}_t^{(k)} }{\mathcal{Z}_t\left(\{p_t,\tilde{p}_t\}\right)}
\end{equation}
which allows for an on-demand exact and rapid computation of the log-infidelity knowing only the $2n$ independently computable bit probabilities.

\begin{figure*}
\centering
\includegraphics[width=0.98\columnwidth,trim = 0.5cm 0cm 0.5cm 0cm, clip]{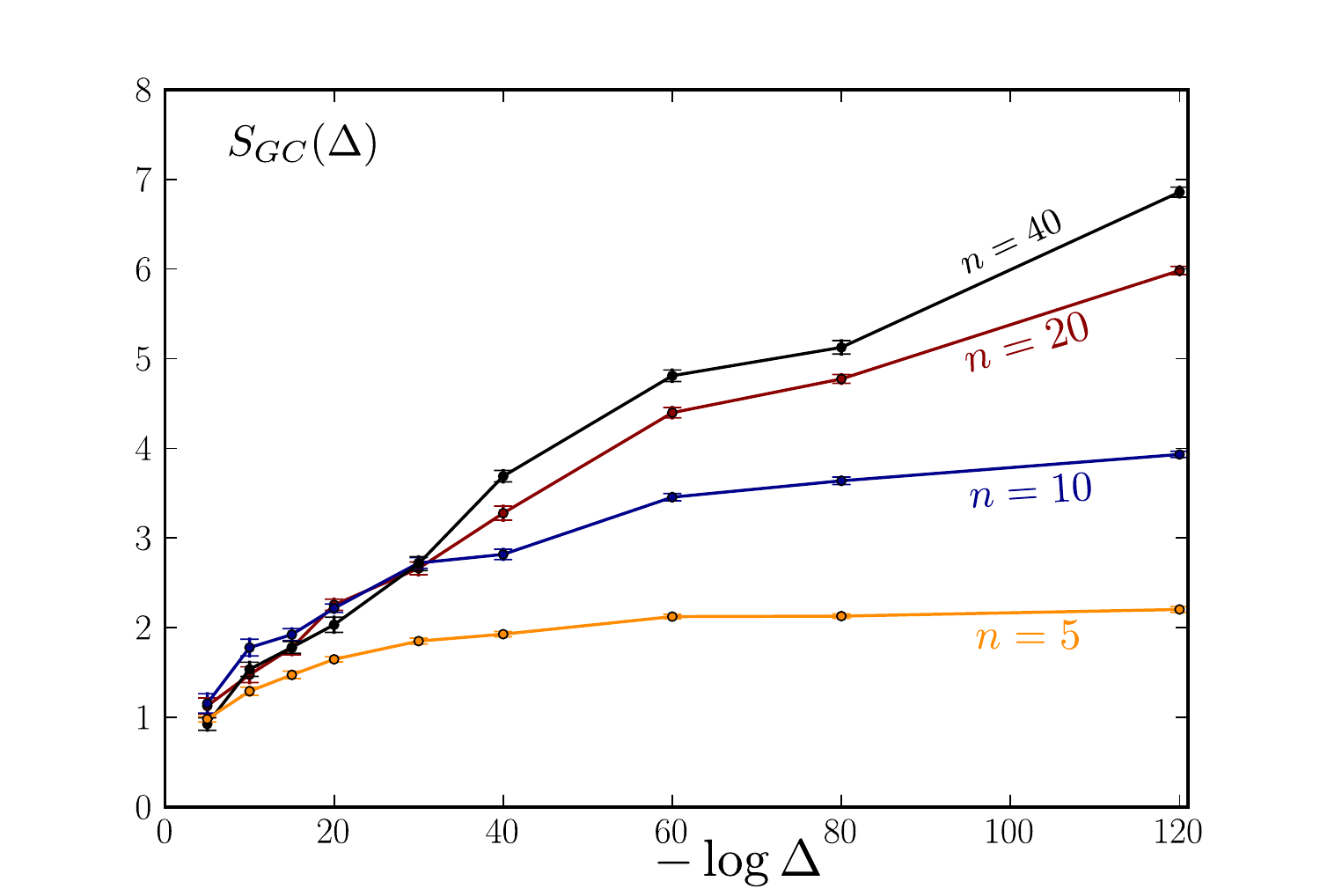}
\includegraphics[width=0.98\columnwidth,trim = 0.5cm 0cm 0.5cm 0cm, clip]{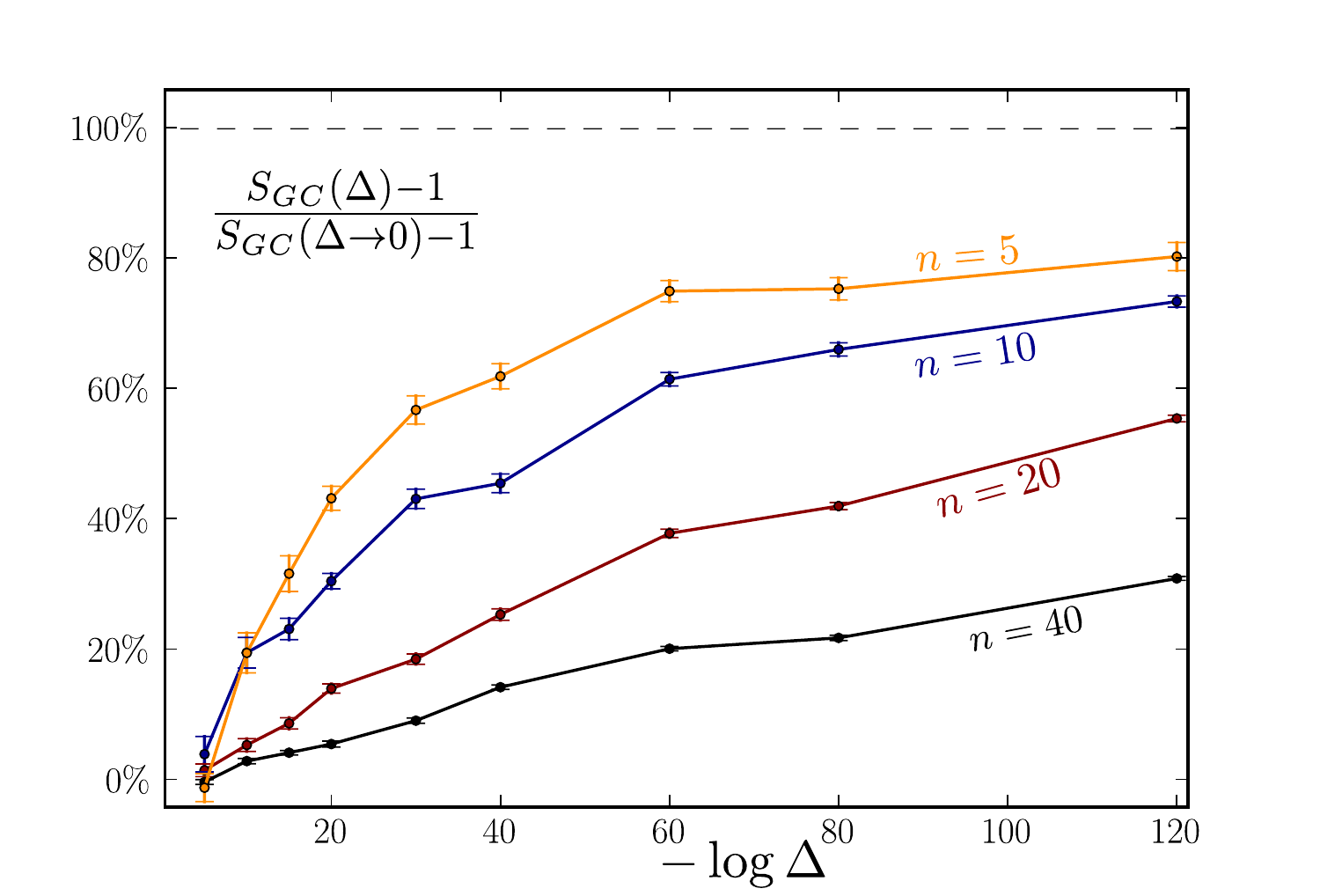}
\caption{On the left, speed-up provided by the GC algorithm for finite infidelity target $\Delta$ for various values of the number of qubits $n$. On the right, percentage of the asymptotic speed-up reached for finite infidelity target. For large values of $n$, the asymptotic speed-up is only approached for extremely small values of $\Delta$. The computations are done with thresholds $p_0=0.5$ and $p'_0=0.001$.}
\label{fig:numerics}
\end{figure*}

\section{Numerics}\label{sec:numerics}

Computing the speed-up rate numerically is useful for two reasons. First, the results previously derived are asymptotic and the speed-up could very well be much smaller for a reasonable non-zero infidelity target. Second and most importantly, computing the speed-up rate numerically for large values of $n$ is the best way to make sure that the protocol does not require an exponential amount of memory and is indeed easily implementable.

The numerical computations can be easily carried out with the help of the discrete equation \eqref{eq:discrete} for $\varepsilon \ll 1$. Starting from a fully unknown register state, the time to reach a given infidelity target is computed for the standard no control scheme and for the GC procedure from which the non-asymptotic speed-up rate is computed. The results for various register sizes are shown in Fig. \ref{fig:numerics}. Unsurprisingly, for large registers, the asymptotic speed-up is only reached for absurdly small infidelity targets. This is because most of the time is spent in the ``guess'' phase trying to find a candidate. Optimizing over the thresholds $p_0$ and $p_0^{'}$ would probably slightly tame this noxious waste of time. Further, the suboptimal step 4 of the procedure does lead to a substantial slow down for reasonable infidelity targets and a less naive procedure might be able to make the non-asymptotic part of the algorithm less costly. 

Alternatively, one could imagine a multi-stage algorithm where a global candidate is found by applying the GC procedure on a series of subregisters, \ie where the ``guess'' phase itself is sped up using the GC algorithm. In any case, these numerical simulations show that the asymptotic speed-up should not be the only metric used to assess the efficiency of rapid measurement schemes in the future as the asymptotic regime may be irrelevant in realistic setups. Note that the previous schemes of Combes \textit{et al.} \cite{combes2008,combes2015} which could only be probed numerically for small registers, also showed lower performances for finite infidelity targets. All these reserves being made, the GC algorithm does provide a large speed-up in absolute value for all register sizes and reasonably small infidelity targets. As a result, and even without the previously mentioned potential improvements, the GC algorithm can be applied profitably to the rapid measurement of qubit registers.

\section{Discussion} \label{sec:discussion}

\begin{figure}
\includegraphics[width=\columnwidth,trim = 0cm 21.5cm 9.5cm 0cm, clip]{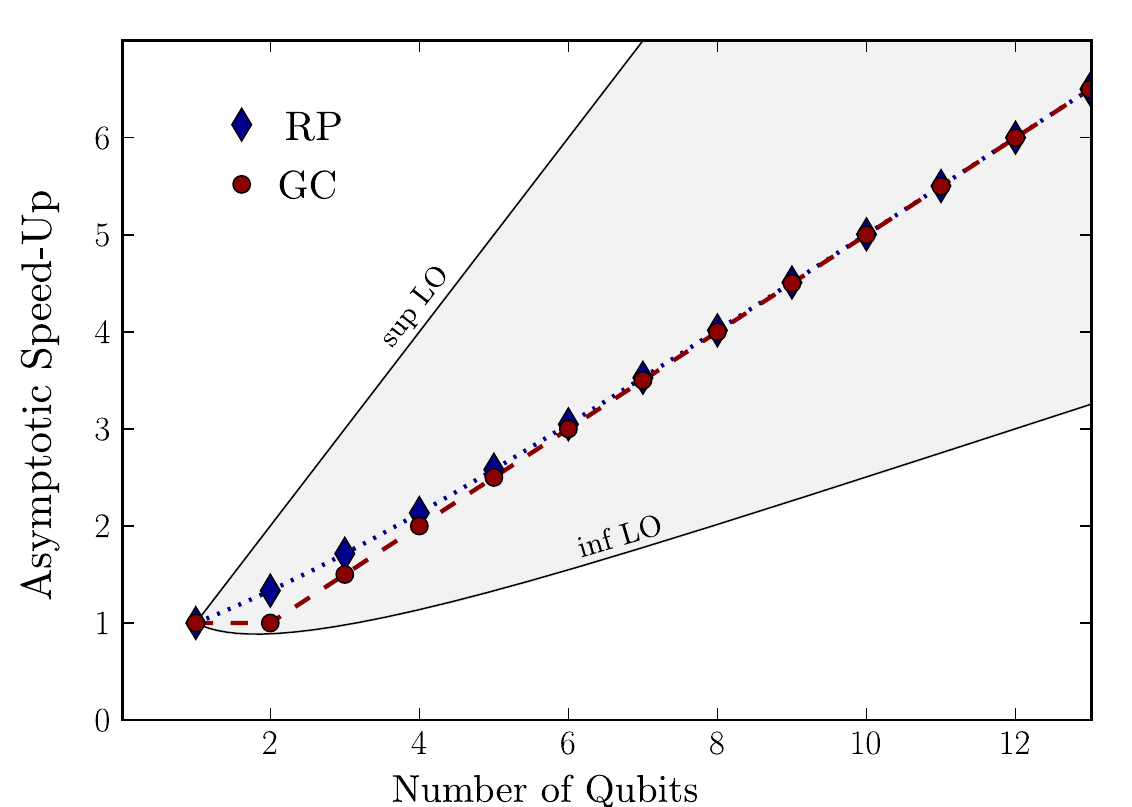}
\caption{Exact asymptotic speed-up for the Random Permutation scheme (RP) and the Guess and Check (GC) scheme computed in this article. The two lines ``sup LO'' and ``inf LO'' show the analytic upper and lower bounds known for the Locally Optimal measurement scheme of \cite{combes2015}.}
\label{fig:comparison}
\end{figure}

We have proposed a new and simple protocol (GC) aimed at increasing the measurement rate of qubit registers and derived the exact asymptotic speed-up it provides. Its asymptotic speed-up rate, compared to that of the earlier schemes of \cite{combes2008} and \cite{combes2015}, is displayed in Fig. \ref{fig:comparison}. The main comparative advantage of the procedure does not reside in its performance increase but in its practical and computational simplicity. Indeed, in terms of control, the GC algorithm only requires a finite number (on average $2$ for thresholds $p_0=1/2$ and $p_0^{'}\ll 1$) of simple permutations on a subspace of dimension $2$. Even if the scheme is not, strictly speaking, open-loop, the fact that the control operations can be done with a delay without performance loss makes it much less demanding than a true real-time feedback control scheme. Eventually and most importantly, the GC algorithm allows to encode the probabilities of all the register configurations in $2n$ real numbers in contrast with the prohibitive exponential number required by other protocols. This last feature makes the GC algorithm, or other algorithms built upon similar ideas, particularly suitable for the rapid measurement of future quantum computer registers where an exponential memory scaling will simply be prohibitive.

Although this was not strictly necessary for our derivation, we have used a classical probabilistic picture throughout the paper. As the rapid measurement problem is essentially a classical problem, recasting everything in an equivalent classical language provides a simpler and hopefully more pedagogical introduction to the subject. Additionally, it helps give intuitive and straightforward answers to questions otherwise non trivial like the sensitivity to control imperfections or the meaning of the result obtained at the end of the protocol. Since the core of the procedure is classical, one may wonder if quantum mechanics nevertheless plays a useful role in its implementation. The answer is positive, the GC scheme requires a permutation of two configurations, something which is highly non intuitive and with a problematic implementation in classical mechanics but which may be carried out with a simple Hamiltonian in quantum mechanics. Consequently, even if the intuition backing the GC protocol is mostly classical, it is likely that it is only implementable on a genuinely quantum system in practice.

In the future, ideas similar to the one developed in this article could be applied to the rapid measurement of more general quantum systems. Even in the restricted context of qubit registers, the numerical simulations have shown that improving the short time behavior of the algorithm could greatly improve its performance in practice. For that matter, analytic studies of the finite time behavior of the log infidelity could certainly be illuminating. 

\begin{acknowledgments}
I thank Denis Bernard and Michel Bauer for useful discussions.
This work was supported in part by the ANR contract ANR-14-CE25-0003-01.
\end{acknowledgments}

\appendix

\section{Exact speed-up for open-loop control}\label{appendix:RP}
The exact speed-up can actually be computed using an exact solution of the stochastic differential equation and a bit of combinatorics. The method is very similar to the one used by Combes \textit{et al.} in \cite{combes2015} to compute the convergence rate of the infidelity in the no feedback case. 

We start by assuming that the random permutations are carried out very quickly so that the whole permutation group is sampled in any infinitesimal time interval. Notice again that this is just needed for simplicity, the order in which the measurements are done does not matter so we can reorganise them a posteriori to fulfill the previous condition. In this setting the register density matrix verifies:
\begin{equation}\label{eq:permutations}
\upd \rho_t =\left[\frac{2\gamma}{(2^n)!}\right]^{1/2} \sum_{\tau\in\mathfrak{S}(2^n)}\sum_{k=1}^{2^n} \mathcal{H}[\sigma^{k,\tau}_z](\rho_t) \upd W_t^{(k,\tau)}
\end{equation}
where $\mathfrak{S}(2^n)$ is the permutation group on the set $\mathscr{S}$ of the $2^n$ configurations, $\sigma_z^{k,\tau}=U_\tau^{-1}\sigma^{k}_z U_\tau$ where $U_\tau$ is the unitary operator implementing the permutation $\tau$ and the $W_t^{(k,\tau)}$ are independent Wiener processes \ie $\upd W_t^{(k,\tau)}\upd W_t^{(k',\tau')}=\delta_{k,k'}\delta_{\tau,\tau'}\upd t$. Equation \eqref{eq:permutations} is invariant under the change $\sigma^{k,\tau}_z\rightarrow\sigma^{k,\tau}_z+\mathds{1}$ which allows to work with projectors on spaces of dimension $2^{n-1}$. There are \emph{only} $\binom{2^n}{\,2^{n-1}\,}$ such projectors which allows the following factorisation.

\begin{equation}
\upd \rho_t = 2 \sqrt{2\gamma\,\mathcal{N}}  \sum_{\mathcal{P}\subset\mathscr{S}, |\mathcal{P}|=2^{n-1}} \mathcal{H}[P^\mathcal{P}](\rho_t) \upd  W_t^{\mathcal{P}}
\end{equation}
with
\begin{equation} 
\mathcal{N}=\frac{n}{\binom{2^n}{\,2^{n-1}\,}}
\end{equation}
and $P^{\mathcal{P}}$ denotes the projector on the subset $\mathcal{P}$ of the set of possible register configuration $\mathscr{S}$. The new Wiener processes are obtained as a normalized sum of the previous independent Wiener processes and are thus also independent Wiener processes.
For pure mathematical convenience we can associate a corresponding set of signals which will allow us to work with what is often called linear quantum trajectories \cite{wiseman2009} (the knowledge of which is not needed here):
\begin{equation}\label{eq:signalbis}
\upd Y^{\mathcal{P}}_t=4\sqrt{2\gamma\,\mathcal{N}}\,\tr\left[P^\mathcal{P} \rho_t\right] \upd t + \upd W^\mathcal{P}_t.
\end{equation}
We introduce an auxiliary density matrix $\tilde{\rho}$ verifying the linear SDE:
\begin{equation}
\upd\tilde{\rho}_t = 4\sqrt{2\gamma\,\mathcal{N}} \sum_\mathcal{P} P^{\mathcal{P}}\tilde{\rho}_t\, \upd Y^\mathcal{P}_t.
\end{equation}
One can verify, using the It\^o formula, that the \emph{true} density matrix $\rho$ can be recovered from $\tilde{\rho}$ through a simple normalisation.
The previous equation can easily be expanded and gives in components:
\begin{equation}
\upd \tilde{\lambda}_s = 4\sqrt{2\gamma\,\mathcal{N}} \,\tilde{\lambda}_s \sum_{\mathcal{P}, s\in \mathcal{P}} \upd Y^\mathcal{P}_t,
\end{equation}
where $\tilde{\lambda_s}$ is the eigenvalue of $\tilde{\rho}$ associated to the state (or configuration) $s$ and which, once normalized, will give its probability.
This equation can be solved exactly as a function of the $Y^\mathcal{P}_t$'s:
\begin{equation}
\tilde{\lambda}_s=\exp\left(4\sqrt{2\gamma\,\mathcal{N}} \,\sum_{\mathcal{P}, s\in \mathcal{P}} Y^\mathcal{P}_t - 8\gamma\,\mathcal{N}\binom{2^n}{\,2^{n-1}\,}\,t\right),
\end{equation}
Finally we can express the normalized probability for the state $s$ of maximum probability:
\begin{equation}
\lambda_0=\frac{\exp\left(4\sqrt{2\gamma\,\mathcal{N}} \,\sum_{\mathcal{P}, 0\in \mathcal{P}} Y^\mathcal{P}_t\right)}{\sum_s \exp\left(4\sqrt{2\gamma\,\mathcal{N}} \,\sum_{\mathcal{P}, s\in \mathcal{P}} Y^\mathcal{P}_t\right)}
\end{equation}
Up to now, everything is exact and some approximations are now needed to work out the large time limit. In this limit $\lambda_0$ is close to one and all the other probabilities are much smaller and decrease exponentially (on average) as a function of time. Using eq. \eqref{eq:signalbis} we thus get:
\begin{equation}
\begin{split}
Y_t^\mathcal{P}&\underset{t+\infty}{\sim} 4\sqrt{2\gamma\,\mathcal{N}}\, t \; \text{if} \;0 \in \mathcal{P}\\
Y_t^\mathcal{Q}&=\underset{t+\infty}{o}(t)  \; \text{if}  \;0 \notin \mathcal{Q}
\end{split}
\end{equation}
Now, one only needs to notice that for the sum over subsets containing $0$ there are $\binom{2^n-1}{2^{n-1}-1}=\binom{2^n}{2^{n-1}}/2$ non negligible terms whereas for the sum over subsets containing $s\neq 0$, there are only $\binom{2^n-2}{2^{n-1}-2}=\frac{1}{4}\frac{2^n-2}{2^n-1}\binom{2^n}{\,2^{n-1}\,}$ non negligible terms:
\begin{equation}
\begin{split}
\sum_{\mathcal{P}, 0\in \mathcal{P}} Y^\mathcal{P}_t& \underset{t+\infty}{\sim}\frac{1}{2}\binom{2^n}{\,2^{n-1}\,} \times 4\sqrt{2\gamma\,\mathcal{N}}\, t \\
\sum_{\mathcal{P}, s\neq0 \in \mathcal{P}} Y^\mathcal{P}_t& \underset{t+\infty}{\sim}\frac{1}{4}\frac{2^n-2}{2^n-1}\binom{2^n}{\,2^{n-1}\,} \times 4\sqrt{2\gamma\,\mathcal{N}}\, t 
\end{split}
\end{equation}
Which gives:
\begin{equation}
\begin{split}
\ln (1-\lambda_0)&\underset{t+\infty}{\sim}-\frac{1}{4}\frac{2^n}{2^n-1} \binom{2^n}{\,2^{n-1}\,} \left(4\sqrt{2\gamma\,\mathcal{N}}\right)^2 t\\
&\underset{t+\infty}{\sim} \frac{n}{2}\frac{2^n}{2^n-1} \times  - 16\, \gamma\,t 
\end{split}
\end{equation}
So that the exact speed-up rate reads:
\begin{equation}
S_{RP} = \frac{n}{2}\,\frac{2^n}{2^n-1} \underset{n+\infty}{\sim} \frac{n}{2}
\end{equation}
which is what we had claimed. This exact result coincides with the upper bound proposed in \cite{combes2015} and seems consistent with its numerical results.

\section{Speed-up rate for \emph{Guess and Check} control}\label{appendix:GC}
In this section, we compute the speed-up rate for the ``Check'' part of the algorithm, assuming the candidate picked at the beginning turns out to be correct, \ie that it stays the most probable state during the whole process. Without lack of generality, we write then $\bar{0}$ the candidate (a notation valid in the two basis). We also assume for convenience that the measurements are made in the two basis simultaneously, which as we argued before, does not change anything in the statistics as long as the same amount of time is spent in each basis. The system density matrix verifies the following SDE: 
\begin{equation}\label{eq:gc}
\begin{split}
\upd \rho_t =\;&\sqrt{\gamma} \sum_{k=1}^n \mathcal{H}[\sigma_z^{(k)}]\left(\rho_t\right)\,\upd W_t^{(k)}\\
+&\sqrt{\gamma}\sum_{k'=1}^n \mathcal{H}[\sigma_z^{(k')}]\left(\tilde{\rho}_t\right)\,\upd \tilde{W}_t^{(k')}
\end{split}
\end{equation}
where $\tilde{\rho}_t$ is the matrix of $\rho_t$ in $\tilde{\mathscr{B}}$ and $W_t^{(k)}$, $\tilde{W}_t^{(k)}$ are independent Wiener processes. As in the previous case, we make the transformation $\sigma_z^{(k)}\rightarrow\sigma_z^{(k)}+\mathds{1}\equiv 2P^{(k)}$ which leaves the previous SDE invariant.
The associated signals $Y^{(i)}$ and $\tilde{Y}^{(i)}$ verify:
\begin{equation}\label{eq:signals}
\begin{split}
\upd Y^{(i)}_t &= 4\sqrt{\gamma} \, \tr \left[P^{(i)} \rho_t\right]\, \upd t + \upd W^{(i)}_t\\
\upd \tilde{Y}^{(i)}_t &= 4\sqrt{\gamma} \, \tr \left[P^{(i)} \tilde{\rho}_t\right]\, \upd t + \upd \tilde{W}^{(i)}_t
\end{split}
\end{equation}
As in the previous appendix, we can solve eq. \eqref{eq:gc} explicitly (as a function of the signal) using a linearised version of the SDE:
\begin{equation}
\upd \bar{\rho}_t=4\sqrt{\gamma} \sum_{k=1}^n\left[ P^{(k)}\bar{\rho}_t \upd Y^{(k)}_t + \tilde{P}^{(k)}\bar{\rho}_t \upd \tilde{Y}^{(k)}_t\right]
\end{equation}
which is solved in components:
\begin{equation}\label{eq:nonnormalised}
\begin{split}
\bar{\lambda}_s=&\exp \left(4\sqrt{\gamma} \sum_{k=1}^n \left[(1-s^{(k)})Y^{(k)} + (1-\tilde{s}^{(k)}) \tilde{Y}^{(k)}\right]\right)\\
&\times \exp\left(-8\gamma \sum_{k=0}^n \left[(1-s^{(k)}) + (1-\tilde{s}^{(k)}\right]\, t\right).
\end{split}
\end{equation}
At large time, when most of the probability is concentrated on the state $s=\bar{0}$, eq. \eqref{eq:signals} gives:
\begin{equation}
\begin{split}
Y^{(k)}_t&\underset{t+\infty}{\sim} 4\sqrt{\gamma} t\\
\tilde{Y}^{(k)}_t&\underset{t+\infty}{\sim} 4\sqrt{\gamma} t
\end{split}
\end{equation}
From eq. \eqref{eq:nonnormalised} it is then easy to see that the non-normalised eigenvalues have three possible behaviors:
\begin{equation}
\begin{split}
\bar{\lambda}_{\bar{0}}&=\exp\left[16n\gamma t + o(t)\right]\\
\bar{\lambda}_{\bar{1}}&=\exp\left[o(t)\right]\\
\bar{\lambda}_s&=\exp\left[8n\gamma t + o(t)\right] \;\;\;\text{for}\;\;s\neq\bar{0}, \bar{1}
\end{split}
\end{equation}
Recalling that 
\begin{equation}
\lambda_{\bar{0}}=\frac{\bar{\lambda}_{\bar{0}}}{\sum_{s\in\mathscr{S}} \bar{\lambda}_s},
\end{equation}
we finally get, for $n\neq1$:
\begin{equation}
\ln(\Delta_t)=\ln (1-\lambda_{\bar{0}})\underset{t+\infty}{\sim}   - 16\, \gamma\,t \times \frac{n}{2} 
\end{equation}

\bibliography{main}

\end{document}